\documentclass[aps,pra,twocolumn]{revtex4}

\usepackage{epsfig,pstricks}

\usepackage{tikz}
\usetikzlibrary{3d,calc}
\usetikzlibrary{arrows}
\usepackage{pgfplots}
\pgfplotsset{compat=1.11}
\usetikzlibrary{calc,fadings,decorations.pathreplacing}
\usetikzlibrary{positioning}
\usepackage{amsmath,amssymb}
\usepackage{color}

\tikzset{%
  >=latex,
  inner sep=0pt,%
  outer sep=2pt,%
  mark coordinate/.style={inner sep=0pt,outer sep=0pt,minimum size=3pt,
    fill=black,circle}%
}

\def\CC{{\rm\kern.24em \vrule width.04em height1.46ex depth-.07ex
\kern-.30em C}}
\def\P{{\rm I\kern-.25em P}}
\def\NN{{\rm I\kern-.25em N}}
\def\RR{{\rm
         \vrule width.04em height1.58ex depth-.0ex
         \kern-.04em R}}
\def\id{{\rm 1\kern-.22em l}}
\def\ZZ{{\sf Z\kern-.44em Z}}
\def\tr{{\rm tr}\;}
\newtheorem{pdef}{Definition}[section]

\newenvironment{eqblock}[2]{\beq\label{#2}\begin{array}{#1}}{\end{array}
                                \eeq}
\newenvironment{neqblock}[1]{\[\begin{array}{#1}}{\end{array}\]}

\newcommand{\ketbra}[1]{\ensuremath{| #1 \rangle \langle #1 |}}

\newcommand{\beqb}{\begin{eqblock}}
\newcommand{\eeqb}{\end{eqblock}} 
\newcommand{\nbeqb}{\begin{neqblock}}
\newcommand{\neeqb}{\end{neqblock}} 

\newcommand{\beq}{\begin{equation}}
\newcommand{\beqa}{\begin{eqnarray}}
\newcommand{\eeq}{\end{equation}}
\newcommand{\eeqa}{\end{eqnarray}}
\newcommand{\nbeqa}{\begin{eqnarray*}}
\newcommand{\neeqa}{\end{eqnarray*}}

\newcommand{\ket}[1]{| #1 \rangle}

\newcommand{\Matrix}[2]{\left( \begin{array}{#1} #2 \end{array}
  \right)}

\def\DJo{$\;$\kern-.4em \hbox{D\kern-.8em\raise.15ex\hbox{--}\kern.35em okovi\'c}}

\begin{document}

\title{Generalized W-state of four qubits with exclusively threetangle}
\author{Sebastian Gartzke and Andreas Osterloh}
\affiliation{Institut f\"ur Theoretische Physik, 
         Universit\"at Duisburg-Essen, D-47048 Duisburg, Germany.}
\email{andreas.osterloh@uni-due.de}
\begin{abstract}
We single out a class of states possessing only threetangle but distributed all over four qubits. 
This is a three-site analogue of states from the $W$-class, which only possess globally distributed 
pairwise entanglement as measured by the concurrence. 
We perform an analysis for four qubits, showing that such a state indeed exists. 
To this end we analyze specific states of four qubits that are not convexly 
balanced as for $SL$ invariant families of entanglement, but only
affinely balanced.
For these states all possible $SL$-invariants vanish, hence they are part of the $SL$ null-cone.  
Instead, they will possess at least a certain unitary invariant.

As an interesting byproduct it is demonstrated that the exact convex roof is reached
in the rank-two case of a homogeneous polynomial $SL$-invariant measure of entanglement 
of degree $2m$, 
if there is a state which corresponds to 
a maximally $m$-fold degenerate solution in the zero-polytope that can be combined 
with the convexified minimal characteristic curve to give a decomposition of $\rho$.  
If more than one such state does exist in the zero polytope, a minimization must be performed.
A better lower bound than the lowest convexified characteristic curve 
is obtained if no decomposition of $\rho$ is obtained in this way.
\end{abstract}

\maketitle

\section{Introduction}

W-states are at the borderline between three distinct and important features 
of multipartite entanglement:
pure W-states are satisfying the 
Coffman-Kundu-Wootters inequality\cite{Coffman00,Osborne06} as an equality\cite{Coffman00},
they appear as representative of one of two classes of entanglement\cite{Duer00} for three qubits, 
and it is seen to be related to 
a ladder from $SL$ invariance to $U$ invariance\cite{JohanssonO13} for an arbitrary 
number of qubits.
Indeed, three qubits have been shown to separate into the GHZ-class which is detected by the 
threetangle\cite{Coffman00} and 
the remaining W-class sharing entanglement among two parties only\cite{Coffman00}. 
A further peculiarity of the $q$ qubit W-state is hence 
that it has no $SL$ invariant $n$-tangle with $n>2$~\cite{OS05,OS09}. 
\\
It is therefor reasonable to ask the following question: do such states also exist for $q$ qubits
and an arbitrary $n<q$;
in other words: are there certain $q$ qubit states possessing only $n$-tangle? 
The corresponding states should 
in particular have no $SL$-invariant, we call it {\it $q$-tangle},  
(with a $(2m,0)$ bidegree of unitary invariants)
and thus be part of the $SL$ null-cone.
The $SL$ null-cone, however, has a finer structure which is classified by further $SU$ invariants 
with a general bidegree $(2m-l,l)$ (see e.g. \cite{Luque06}).
$SU$-invariants of bidegree $(2m-l,l)$ are $(2m-l)$-linear in the wave-function $\psi$
and $l$-linear in its complex conjugate $\psi^*$ (or vice-versa). 
Every state which lies outside the null-cone 
must futhermore have a part which is balanced\cite{OS09} or equivalently termed c-balanced (c for convex) 
in Ref.~\cite{JohanssonO13}.
In contrast, there are those states which are a-balanced (a for affine) without being c-balanced. 
These states have been singled out having discrete topological phases under 
the cyclic local $SU$ group-operation\cite{Johansson12}
and emerge from the c-balanced states by means of {\em partial spin flips}~\cite{JohanssonO13}.
Simple examples are the states in the $SL$ W-class 
\beq
\ket{\psi_{\rm{W-class}}}=a_0\ket{000}+a_1\ket{100}+a_2\ket{010}+a_3\ket{001}
\eeq 
for three qubits which do possess a $SU$ $(3,1)$-invariant.
These states, by means of a partial spin flip, are connected to the $(4,0)$-invariant 
states which are in the $SL$-invariant $GHZ$-class 
\beq
\ket{\psi_{\rm{GHZ-class}}}=a_0\ket{111}+a_1\ket{100}+a_2\ket{010}+a_3\ket{001}\ .
\eeq
In both formulae $a_i\neq 0$ for $i=0,\dots,3$.
The original W-states are however $(q,q)$-invariant and
are $SL$- but not $U$-equivalent to $\ket{\psi_{\rm{W-class}}}$. 
They do not emerge from this procedure after performing $q$ partial spin flips
since they are completely unbalanced states\cite{OS09}. They are however obtained, when omitting some 
product basis state from the outcome of such a procedure.
Nevertheless, every state displaying a unitary $(2m-l,l)$-invariant and 
which therefor has no $(2m,0)$-invariant will be 
a good starting point to look at as soon as it is not bipartite.

The work is organized as follows: in the next section we descibe the states we are analysing.
In section~\ref{convex-roof} we emphasize on details about the calculation of the convex-roof 
singling out those states which merely contain threetangle. In the conclusions we discuss
the obtained results and give an outlook.  

\section{States from the $SL$ null-cone}\label{states}

We start from the four-qubit maximally entangled c-balanced states 
\beqa
\ket{\Psi^4_6}&=&\frac{1}{\sqrt{3}}\ket{1111}+\sqrt{\frac{2}{3}}\ket{W^4}\label{6er}\\
\ket{\Psi^4_4}&=&\frac{1}{2}(\ket{1111}+\ket{1100}+\ket{0010}+\ket{0001})\label{4er}
\eeqa
Here, $\ket{\Psi^q_{L}}$ means a $q$ qubit state which is irreducibly c-balanced 
of length $L=2n$\cite{OS09}, and $\ket{W^4}=(\ket{1000}+\ket{0100}+\ket{0010}+\ket{0001})/2$.

\subsection{States derived from $\Psi^4_6$}

The state taken from Eq.~\eqref{6er} is detected by the only 
genuine $(6,0)$-filter-invariant\cite{OS04,DoOs08,OS09,JohanssonO13} 
of $SU$ giving a non-zero result due to its length. 
Possible states in the $SL$ null-cone therefore
have $(5,1)$-, $(4,2)$-, and $(3,3)$-invariance\cite{JohanssonO13} and are obtained by a partial 
spin flip on one, two, or three components respectively of the product basis. 
Since the state is translation symmetric (even with respect to the symmetric group of permutations)
it does not matter on which of the four components of the $W^4$-state the partial spin flips 
are acting on. Therefore we have a single case of $(5,1)$-, $(4,2)$-, and $(3,3)$-invariance each
and one $(4,2)$-invariant acting on the $\ket{1111}$-component together with a $(3,3)$-invariant 
if the next partial spin flip is acting on the $W^4$-state. The $(3,3)$-invariant states, 
however, are bipartite product states and therefore will not be considered any further.
This translates into the following states
\beqa
\ket{\Psi^4_{6;1}}&=&\frac{1}{\sqrt{3}}\ket{0000}+\sqrt{\frac{2}{3}}\ket{W^4}\\
\ket{\Psi^4_{6;2}}&=&\frac{1}{\sqrt{3}}\ket{1111}+\nonumber\\
&&\ \frac{1}{\sqrt{6}}(\ket{0111}+\ket{0100}+\ket{0010}+\ket{0001})\\
\ket{\Psi^4_{6;23}}&=&\frac{1}{\sqrt{3}}\ket{1111}+\nonumber\\
&&\ \frac{1}{\sqrt{6}}(\ket{0111}+\ket{1011}+\ket{0010}+\ket{0001})
\eeqa
where the indices after the colon in 
$\ket{\Psi^4_{6;234}}$ would indicate the partial spin flip operation, acting here on the components 
$2$, $3$, and $4$ of the state $\ket{\Psi^4_6}$.

Whereas the state $\ket{\Psi^4_{6;1}}$ becomes a mixture of states in the W-class and 
therefore contains no threetangle, 
$\ket{\Psi^4_{6;2}}$ and $\ket{\Psi^4_{6;23}}$ may contain threetangle instead.

\subsection{States derived from $\Psi^4_4$}

The state taken from Eq.~\eqref{4er} is detected by all of the three $(4,0)$-invariants 
which are called ${\cal C}^{(4)}_{ij}$ in Ref.~\cite{DoOs08} 
respectively ${\cal B}^{I}_{[4]}$, ${\cal B}^{II}_{[4]}$, and
${\cal B}^{III}_{[4]}$ in Ref.~\cite{Viehmann}. It is a state which has length $4$ and hence 
cannot be detected by the $(6,0)$-filter invariant as the original state considered previously.
Due to the symmetries of the state by permutations of the qubits 
there are only four distinct states in the $SL$ null-cone:
three states have a $(3,1)$-invariant; 
the one with a $(2,2)$-invariant of $SU$ is a bipartite state 
and will therefore not be considered.
These three states are:
\beqa\label{state:4-1}
\ket{\Psi^4_{4;1}}&=&\frac{1}{2}(\ket{0000}+\ket{1100}+\ket{0010}+\ket{0001})\\
\ket{\Psi^4_{4;2}}&=&\frac{1}{2}(\ket{1111}+\ket{0011}+\ket{0010}+\ket{0001})\\
\ket{\Psi^4_{4;4}}&=&\frac{1}{2}(\ket{1111}+\ket{1100}+\ket{0010}+\ket{1110})\label{state:4-2}
\eeqa
The notations is reflecting where the partial spin flip is acting on as in the previous section.
Here however the trace over the first or second qubit leads to a mixed state that
is free from threetangle whereas tracing out one of the other two sites 
leads to a reduced density matrix that may contain threetangle.

\section{Convex-roof construction}\label{convex-roof}
Since we intend to find a state with threetangle distributed all over the chain
and ideally without any concurrence, we look at first to the reduced three-site density matrices.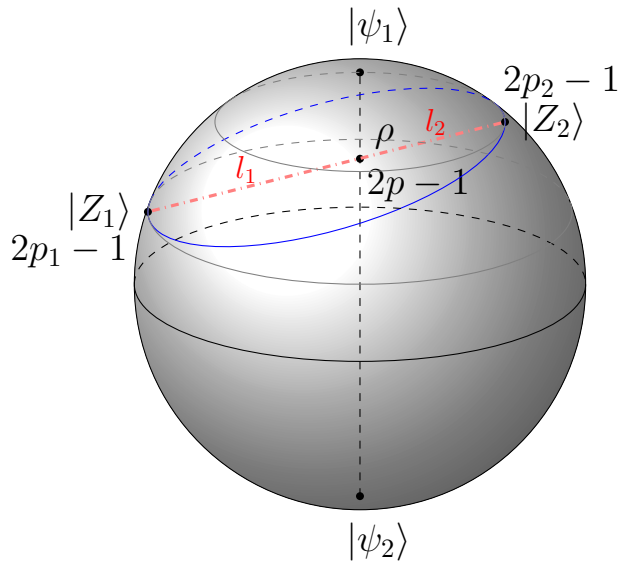
\begin{figure}
\centering
\begin{tikzpicture}[invclip/.style={insert path={(-5,-5) rectangle (5,5)}}]
\newcommand\pgfmathsinandcos[3]{%
  \pgfmathsetmacro#1{sin(#3)}%
  \pgfmathsetmacro#2{cos(#3)}%
}
\newcommand\LongitudePlane[3][current plane]{%
  \pgfmathsinandcos\sinEl\cosEl{#2} 
  \pgfmathsinandcos\sint\cost{#3} 
  \tikzset{#1/.style={cm={\cost,\sint*\sinEl,0,\cosEl,(0,0)}}}
}
\newcommand\LatitudePlane[3][current plane]{%
  \pgfmathsinandcos\sinEl\cosEl{#2} 
  \pgfmathsinandcos\sint\cost{#3} 
  \pgfmathsetmacro\yshift{\cosEl*\sint}
  \tikzset{#1/.style={cm={\cost,0,0,\cost*\sinEl,(0,\yshift)}}} %
}
\newcommand\CoordPlane[4][current plane]{%
  \pgfmathsinandcos\sinEl\cosEl{#2} %
  \pgfmathsinandcos\sint\cost{#3} %
  \pgfmathsinandcos\sinphi\cosphi{#4} %
  \tikzset{#1/.style={cm={\cost*\cosphi*\cost,-\sinphi*\sint*\sinEl,0,\sint*\cosphi*\cosEl,(0,\yshift)}}}
}
\def\R{3} 
\def\angEl{20} 
\def\angPhia{90} %
\def\angThea{40} %
\def\angPhib{270} %
\def\angTheb{70} %
\pgfmathsetmacro{\H}{\R*cos(\angEl)} 
\pgfmathsetmacro{\exa}{\R*sin(\angThea)*sin(\angPhia)} 
\pgfmathsetmacro{\yya}{-\R*(sin(\angEl)*cos(\angPhia)*sin(\angThea)-cos(\angEl)*cos(\angThea))} 
\pgfmathsetmacro{\exb}{\R*sin(\angTheb)*sin(\angPhib)} 
\pgfmathsetmacro{\yyb}{-\R*(sin(\angEl)*cos(\angPhib)*sin(\angTheb)-cos(\angEl)*cos(\angTheb)} 
\fill[draw,ball color=white,opacity=50] (0,0) circle (\R); 
\coordinate (O) at (0,0);
\coordinate[mark coordinate] (N) at (0,\H);
\coordinate[mark coordinate] (S) at (0,-\H);
\coordinate[mark coordinate] (Z1) at (\exa,\yya);
\coordinate[mark coordinate] (Z2) at (\exb,\yyb);
\draw[dashed] (0,-\H) -- (0,\H);
\coordinate[mark coordinate] (p0) at (0,1.67);
\node[right=3pt,above=2pt] at (p0) {\qquad\Large $\rho$};
\node[right=3pt,below=2pt] at (p0) {\hspace*{9mm}\qquad\Large $2p-1$};
\draw[dashdotted,very thick,red!50,opacity=20] (Z1) -- (Z2);
\newcommand\DrawLatitudeCircle[3][1]{
  \LatitudePlane{\angEl}{#2}
  \tikzset{current plane/.prefix style={scale=#1}}
  \pgfmathsetmacro\sinVis{sin(#2)/cos(#2)*sin(\angEl)/cos(\angEl)}
  \pgfmathsetmacro\angVis{asin(min(1,max(\sinVis,-1)))}
  \draw[current plane,thin,#3] (\angVis:1) arc (\angVis:-\angVis-180:1);
\begin{scope}
  \clip (\exa,\yya) -- (\exb,\yyb) -- cycle;
  \draw[current plane,thin,dashed,#3!40] (180-\angVis:1) arc (180-\angVis:\angVis:1);
\end{scope}
\begin{pgfinterruptboundingbox} 
\path [clip] (\exa,\yya) -- (\exb,\yyb) -- cycle [invclip];
\end{pgfinterruptboundingbox}
\draw[current plane,thin,#3,dashed] (180-\angVis:1) arc (180-\angVis:\angVis:1);
}
\newcommand\DrawCoordCircle[4][1]{
  \CoordPlane{\angEl}{#2}{#4}
  \tikzset{current plane/.prefix style={scale=#1}}
  \pgfmathsetmacro\sinVis{sin(#2)/cos(#2)*sin(\angEl)/cos(\angEl)}
  \pgfmathsetmacro\angVis{asin(min(1,max(\sinVis,-1)))}
  \draw[current plane,thin,#3] (\angVis:1) arc (\angVis:-\angVis-180:1);
\begin{scope}
  \clip (\exa,\yya) -- (\exb,\yyb) -- cycle;
  \draw[current plane,thin,dashed,#3!40] (180-\angVis:1) arc (180-\angVis:\angVis:1);
\end{scope}
\begin{pgfinterruptboundingbox} 
\path [clip] (\exa,\yya) -- (\exb,\yyb) -- cycle [invclip];
\end{pgfinterruptboundingbox}
\draw[current plane,thin,dashed] (180-\angVis:1) arc (180-\angVis:\angVis:1);
}
\begin{scope}[rotate around={16:(0,0)}]
    \DrawLatitudeCircle[\R]{35}{blue}
\end{scope}
\DrawLatitudeCircle[\R]{0}{black} 
\DrawLatitudeCircle[\R]{90-\angThea}{gray} 
\DrawLatitudeCircle[\R]{90-\angTheb}{gray} 
\node[above=8pt] at (N) {\Large $\quad \ket{\psi_1}$};
\node[below=8pt] at (S) {\Large $\quad \ket{\psi_2}$};
\node[right=4pt] at (Z1) {\Large $\ket{Z_2}$};
\node[left=4pt] at (Z2) {\Large $\ket{Z_1}$};
\node[above=7pt] at (Z1) {\hspace*{15mm}\Large $2p_2-1$};
\node[below=7pt] at (Z2) {\hspace*{-21mm}\Large $2p_1-1$};
\node[above=9pt] at (Z2) {\red\large $\qquad\qquad\qquad\ l_1$};
\node[left=20pt] at (Z1) {\red\large $l_2$};
\end{tikzpicture}
\caption{The Bloch sphere of density matrices made of the orthonormal states
$\ket{\psi_i}$, $i\in\{1,2\}$ is shown, together with the two superposition
states $\ket{Z_i}$, $i\in\{1,2\}$ which decompose the density matrix $\rho$. }\label{Bloch}
\end{figure}

\subsection{Relevant formulae}

For completeness we give explicit formulae for where the central line
connecting a pure state $\ketbra{Z_1}$ at $2p_1-1$ with the density matrix
\beq
\rho=p\ketbra{\psi_1}+(1-p)\ketbra{\psi_2}
\eeq
hits the surface of the bloch sphere, hence in the pure state $\ketbra{Z_2}$
at $2p_2-1$, which is split into the corresponding lengths $l_1$ and $l_2$ 
according to
\beqa
l_1&=&\sqrt{1+(2p-1)^2-2(2p-1)(2p_1-1)}\; ,\\
l_2&=&\frac{2\sqrt{2}p(1-p)}{\sqrt{1+(2p-1)^2-2(2p-1)(2p_1-1)}}\; ,\\
p_2&=&\frac{p^2(1-p_1)}{p(p-p_1)+p_1(1-p)}\; .
\eeqa
The lengths $l_i$, $i\in\{1,2\}$, are therefor yielding the corresponding weights 
\beq
q_1=\frac{l_2}{l_1+l_2}\quad ;\quad\; q_2=\frac{l_1}{l_1+l_2}
\eeq
that convexly combine the states $\ketbra{Z_i}$, $i\in\{1,2\}$, to $\rho$
(see Fig.~\ref{Bloch}). 

\subsection{States derived from $\Psi^4_6$}
As already mentioned, the state $\ket{\Psi^4_{6;1}}$ possesses, similar to the W-states,
merely concurrence and no threetangle. These states do occur for every number of qubits.
We will term all those states to be of the W-type,
in this case for four qubits, and don't discuss this state any further.

We have two states remaining: a) $\ket{\Psi^4_{6;2}}$, and b) $\ket{\Psi^4_{6;23}}$.

\subsubsection{The state $\ket{\Psi^4_{6;2}}$}

There are only two essentially different cases due to the form-invariance of 
\beqa\label{Psi6:2}
\ket{\Psi^4_{6;2}}&=&\sqrt{p_1}\ket{1111}+\sqrt{p_2}e^{i\eta}\ket{0111}+c_3\ket{0100}\nonumber\\
&&+c_4\ket{0010}+c_5\ket{0001}
\eeqa
with respect to permutations of the last three qubits. The coefficients $c_i\in\CC$, $i=2,\dots,5$,
and the state is normalized: $|c_i|^2=p_i$ with $\sum_{i=1}^5p_i=1$.
This leads to two different reduced classes of three-site density matrices to be considered:
\beqa\label{1stcase}
\tr_1 \ketbra{\Psi^4_{6;2}}\!\!&=&\!\! p_1\ketbra{111} + (\sqrt{p_2}e^{i\eta}\ket{111}+\\
c_3\ket{100}\!\!&+&\!\!c_4\ket{010}+c_5\ket{001})(h.c.)\nonumber\\
\tr_2\ketbra{\Psi^4_{6;2}}\!\!&=& \!\!(\sqrt{p_1}\ket{111} + \sqrt{p_2}e^{i\eta}\ket{011}+\label{2ndcase}
\\
c_3\ket{000})(h.\!\!\!&c.&\!\!\!)\;+\;(c_4\ket{010}+c_5\ket{001})(h.c.)\nonumber 
\eeqa
with $h.c.$ indicating the hermitean conjugation.
Whereas in the second case both states are already orthogonal, we have to do a bit of algebra 
in order to construct the eigenstates for the first instance.

Diagonalizing this density matrices and re-purifying the result, or equivalently, 
applying a proper local unitary 
\beq\label{unitary}
U=\Matrix{cc}{\cos \alpha & e^{i\chi}\sin\alpha \\
-e^{-i\chi}\sin\alpha & \cos\alpha}
\eeq
with the angle $\alpha$ and the phase $\chi$ on the first site leads to
\beqa
\ket{\widetilde{\Psi}^4_{6;2}}&=&\sqrt{p_1(q)}\ket{1111}+\sqrt{p_2(q)}e^{i \eta}\ket{0111}+\\
&&\!\!c_3\ket{0100}+c_4\ket{0010}+c_5\ket{0001}\; .\nonumber
\eeqa
Parametrizing $(2q-1)=:\sin \beta$ and $p_{\rm rem}:=1-\sum_{i=3}^5p_i\;\in\; [0,1]$ we obtain
\beqa
p_1(q)&:=&p_{\rm rem}\cos^2{\beta}=4 p_{\rm rem}q(1-q) \\
p_2(q)&:=&p_{\rm rem} \sin^2{\beta}=p_{\rm rem}(2q-1)^2 \; .
\eeqa
and the condition for $\alpha$ which derives from the orthogonality relation of the two eigenstates is
\beqa\label{def-Q}
p_{\rm rem}\sin(2\beta)&=&\tan(2\alpha)\\
\chi&=&\eta
\eeqa
with the solution for $\alpha$ given by 
\beq
\alpha=\frac{1}{2}{\rm arctan} 
\left[p_{\rm rem}\sin(2\beta)\right]\; .
\eeq
The corresponding eigenstates of the reduced density matrix are
\beqa
\ket{\psi_1}&\!\!=&\!\!\sqrt{p_{\rm rem}}e^{i\eta}\cos{(\alpha+\beta)}\ket{111}
 -\nonumber\\
&&\sin\alpha\ket{W_3(c_3,c_4,c_5)}\\
\ket{\psi_2}&\!\!=&\!\!\sqrt{p_{\rm rem}}e^{i\eta}\sin{(\alpha+\beta)}\ket{111} 
 +\nonumber\\
&&\cos\alpha\ket{W_3(c_3,c_4,c_5)}\; .
\eeqa
The states are normalized to the relative probability with which they occur
in the density matrix, hence
\beq\label{rho-1st}
\rho_1(p_{\rm rem},q)=\tr_1 \ketbra{\Psi^4_{6;2}}=\ketbra{\psi_1}+\ketbra{\psi_2}\; .
\eeq
The corresponding probabilities are the modulus squared of the wavefunctions, i.e.
\beqa
P_1&=&1-p_{\rm rem}\sin^2(\alpha+\beta)\\
P_2&=&1-p_{\rm rem}\cos^2(\alpha+\beta).
\eeqa
\begin{figure}
  \centering
  \includegraphics[width=.95\linewidth]{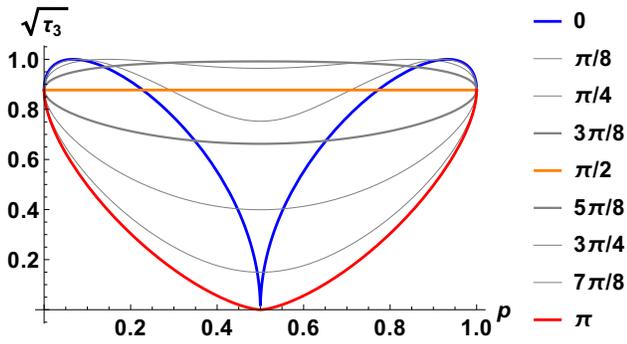}
  \caption{Characteristic curves for certain values of $\varphi=0$ (blue curve that interconnects 
concavely the points at $p=0,1$ with $(p,\sqrt{\tau_3})=(1/2,0) $) to the straight orange line at 
$\varphi=\pi/2$ up to $\varphi=\pi$ (the lowest red curve). 
The curves are symmetrically distributed around $\varphi=0$ and $\varphi=\pi$.
The red lowest curve is the minimal characteristic curve and already convex.
Therfore it constitutes a lower bound to $\widehat{\sqrt{\tau_3}}$.
The zero polytope consists of a threefold degenerate zero at the angle $\varphi=\pi$
and a single zero at $\varphi=0$.}
  \label{charcurves}
\end{figure}
The pure states under consideration are hence
\beq
\ket{\Psi(p,\varphi)}:=\sqrt{pP_1}\ket{\psi_1}-\sqrt{(1-p)P_2}e^{i \varphi}\ket{\psi_2}\; .
\eeq

In what follows, 
we will only consider the case $p_3=p_4=p_5=1/6$. 
This case implies $p_{rem}=1/2$ and the missing probabilities are
$p_1(q)=(\cos^2 \beta)/2=2q(1-q)$ and $p_2(q)=(\sin^2 \beta)/2=(2q-1)^2/2$. Therefore the  
state $\Psi^4_{6;2}$ appears at the value of $1-q=(3+\sqrt{3})/6$ in the following diagrams.
We obtain for the angle for this specific case $\alpha=-{\rm arctan}[\frac{\sqrt{2}}{3}]/2
\approx -0.220255$.

Other values for the probabilities
can be achieved by local $SL$-operations taking into consideration the invariance 
of the corresponding threetangle with respect to these operations. 
It must however be taken care 
that the normalization in general is not conserved.

The characteristic curves, hence the values of their absolute value for
$\sqrt{\tau_3}$, are shown in Fig.~\ref{charcurves} for various values of $\varphi$.
We refer to Ref.~\cite{OstSimplify16} to elucidate the procedure.
\begin{figure}
  \centering
  \includegraphics[width=.95\linewidth]{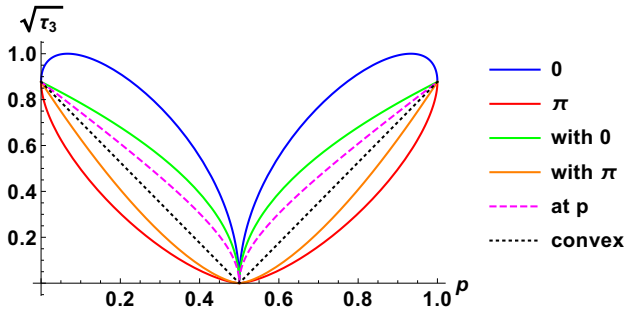}
  \caption{Upper bounds to the convex roof $\widehat{\sqrt{\tau_3}}$ of $\sqrt{\tau_3}$.
The characteristic curves at the angles $\varphi=0,\pi$ are shown together with 
valid decompositions of the density matrix, namely a convex combination of 
a) one of the two eigenstates and 
$\rho_0=\frac{1}{2}(\ketbra{\Psi(\frac{1}{2},0)}+\ketbra{\Psi(\frac{1}{2},\pi)})$
(black dotted lines)
of the states taken from the zero-polytope $\Psi(\frac{1}{2},0)$ and $\Psi(\frac{1}{2},\pi)$,
b) $\ketbra{\Psi(p,0)}$ and $\ketbra{\Psi(p,\pi)}$ (magenta dashed line),
c) the state $\ketbra{\Psi(\frac{1}{2},0)}$ from the zero polytope and a state $\ketbra{\Psi(q,\pi)}$
such that the line connecting both states intersects the center line of the bloch sphere at $(2p-1)$ 
(orange curve below the dotted line),
d) the same as in c) but with $\ketbra{\Psi(\frac{1}{2},\pi)}$ from the zero-polytope
and the corresponding state $\ketbra{\Psi(q,0)}$ (green curve above the dashed and dotted line).
The orange curve coincides with the convex roof.}
  \label{convex-roof}
\end{figure}
Valid decompositions of the density matrix i.e. 
upper bounds to the convex roof $\widehat{\sqrt{\tau_3}}$ of $\sqrt{\tau_3}$
are visualized in Fig.~\eqref{convex-roof}.
They show various convex combinations of $\rho$.
The orange curve is given by the expression
\beq
\widehat{\sqrt{\tau_3}}[\rho_1(\frac{1}{2},p)]=\frac{2}{3^{3/4}}|2p-1|^{3/2}\; ;
\eeq 
it coincides with the convex roof which we explain in what follows.
For $p_0=(3\pm\sqrt{3})/6$ the convex-roof of $\rho_0:=\rho_1(1/2,p_0)$ is 
$\widehat{\sqrt{\tau_3}}[\rho_0]=2/(3\sqrt{3})\approx 0.3849$.

As seen in Fig.~\ref{charcurves}, the characteristic curves are strictly concave around 
the single zero at $\varphi=0$ and $p=1/2$. This has two effects: 
a) any deviation of a decomposition-state around this value is greater than zero
with a square-root behavior. 
And b) the weight of the states appears enhanced if more than one state is comprising the decomposition
yielding a mixed state.
For these two reasons one of the decomposition states is known 
to be the pure state which lies in the minimum of the single zero of the zero-polytope 
with concave behavior (linear behavior is included).
One valid decomposition made of this decomposition state
is the orange curve in Fig.~\ref{convex-roof}. This acquires 
the absolute minimum because the characteristic curve at angle $\varphi=\pi$ is
the minimal characteristic curve which is convex. Hence, any decomposition of pure states 
gives a resulting tangle which lies above that curve, similar to the argument in~\cite{LOSU}.
More generally this is true for any homogeneous polynomial $SL$-invariant measure 
of entanglement of degree $2m$,
if there exists a state which corresponds to a maximally $m$-fold degenerate solution 
in the zero-polytope that can be combined 
with the convexified minimal characteristic curve to give a decomposition of $\rho$, 
as is the case here. 
For a given density matrix $\rho$ the interconnecting straight lines with 
the minimal convexified characteristic curve hits the surface of the zero-polytope
somewhere. A minimization of the effective entanglement measure along this decomposition of 
$\rho$ will give the convex roof. Due to convexity, it is enough minimizing along the curve 
on the surface of the zero-polytope that is facing the minimal characteristic curve
on the bloch sphere.

It must be emphasized that also one of the upper bounds in Ref.~\cite{OstNineClasses16}
was similar to this type except that the lowest convex curve was not convex; 
one should of course consider its convexification for reaching the convex roof.
This should be reconsidered in the future.

If the zeros in the zero-polytope are not precisely at opposite angles of the sphere
the optimal decomposition in the convex roof will change continuously from this 
absolutely optimal decomposition that we have in this case. It can therefor be considered 
as lower bound for this type of solutions and gives a better lower 
bound than the minimal characteristic curve as used in 
Refs.~\cite{Eltschka2012,Siewert2012,Eltschka2014,SiewertPPT2Qutrits} 
to lower bound the convex-roof making use of the symmetry in certain states.

We now briefly come back to the second case, in which the eigenstates can be directly read off
Eq.~\eqref{2ndcase}. 
We have
\beq
\rho_2(\{p_i\};p)=p P_1\ketbra{\psi_1}+(1-p)P_2\ketbra{\psi_2}
\eeq
where
\beqa
\ket{\psi_1}&=& \sqrt{p_1}\ket{111}+\sqrt{p_2}e^{i\eta}\ket{011}+c_3\ket{000}\; ,\\
\ket{\psi_2}&=& c_4\ket{010}+c_5\ket{001}\; ,
\eeqa
and
\beqa
P_1&=&p_1+p_2+p_3\\
P_2&=&p_4+p_5\; .
\eeqa
There is only one convex characteristic curve, 
which is the straight line connecting
zero with $\sqrt{p_1p_3}/(p_1+p_2+p_3)$. Hence
\beq
\widehat{\sqrt{\tau_3}}[\rho_2(\{p_i\};p)]=\frac{p \sqrt{p_1p_3}}{P_1}\; ,
\eeq
which means inserting $p=P_1$ for $\tr_2\ketbra{\Psi^4_{6;2}}$.
For the same choice of probabilities as above we get
$\widehat{\sqrt{\tau_3}}[\rho_2]=\frac{1}{3\sqrt{2}}\approx 0.2357$.

At the end, we briefly comment on the CKW inequality noting that for this particular state
all concurrences do vanish. The extended inequality would however already be satisfied 
with $(\widehat{\sqrt{\tau_3}})^2$ as 
threetangle\cite{AdessoRegula14,AdessoRegula14b,AdessoOsterlohRegula16}.

To conclude, we have found states that, similar to the $W$-class which have 
all of their entanglement stored among two parties distributed all over the chain, 
do possess merely threetangle, again distributed all over the four parties.
The states have the form $\ket{\Psi^4_{6;2}}$ (see Eq.~\eqref{Psi6:2}).

\subsubsection{The state $\ket{\Psi^4_{6;23}}$}

The state 
\nbeqa\label{Psi6:23}
\ket{\Psi^4_{6;23}}&:=&\sqrt{p_1}\ket{1111}+\\
&&\hspace*{-17mm}\sqrt{p_2}e^{i\eta}\ket{0111}+c_3\ket{1011}+c_4\ket{0010}+c_5\ket{0001}
\neeqa 
with $c_i\in\CC$, $i=2,\dots,5$, $|c_i|^2=p_i$, is normalized for $\sum_{i=1}^5 p_i=1$. 
It is form-invariant with respect to permutation of 
the first and the last two qubits. 
Hence, there are also two essentially different reduced
density matrices to be considered. They are
\beqa
\tr_1 \ketbra{\Psi^4_{6;23}}&=& (\sqrt{p_1}\ket{111}+c_3\ket{011})({\rm h.c.})\\
&& \hspace*{-15mm}+ (\sqrt{p_2}e^{i\eta}\ket{111}+c_4\ket{010}+c_5\ket{001})({\rm h.c.})\nonumber\\
\tr_3  \ketbra{\Psi^4_{6;23}}&=& p_5\ketbra{001}\\
&&\hspace*{-35mm}+(\sqrt{p_1}\ket{111}+\sqrt{p_2}e^{i\eta}\ket{011}+c_3\ket{101}+c_4\ket{000})({\rm h.c.})\nonumber
\eeqa
The first reduced density matrix is written in subnormalized eigenvector form
\beq
\rho=\ketbra{\psi_1}+\ketbra{\psi_2}
\eeq
whose subnormalized eigenvectors (obtained with the same method as
in the preceding section) are
\beqa
\ket{\psi_1}&=&\left(\sqrt{p_1}\cos\alpha-\sqrt{p_2}\sin\alpha\right)e^{i\eta}\ket{111}\\
&&+c_3 e^{i\eta}\cos\alpha\ket{011}-\sin \alpha(c_4\ket{010}+c_5\ket{001})\nonumber\\
\ket{\psi_2}&=&\left(\sqrt{p_1}\sin\alpha+\sqrt{p_2}\cos\alpha\right)e^{i\eta}\ket{111}\\
&&+c_3 e^{i\eta}\sin\alpha\ket{011}+\cos \alpha(c_4\ket{010}+c_5\ket{001}) \nonumber
\eeqa
where 
\beqa
\tan 2\alpha &=&\frac{\sqrt{2p_1p_2}}{2(p_1+p_3)-1}\\
\chi&=&\eta
\eeqa
and whose threetangle vanishes.

Only the second reduced density matrix 
\beq
\rho_3=\ketbra{\psi_1}+\ketbra{\psi_2}
\eeq
with
\beqa
\ket{\psi_1}&=&\sqrt{p_1}\ket{111}+c_4\ket{000}+\\
&&\qquad\sqrt{p_2}e^{i\eta}\ket{011}+c_3\ket{101}\nonumber\\
\ket{\psi_2}&=&\sqrt{p_5}\ket{001}
\eeqa
has nontrivial threetangle.
Its zero-simplex consists of a single point at the end of the interval $[0,1]$
which goes back to a fourfold-degenerate root.
It leads consequently to a single linear characteristic curve.
Therefore the convex roof of
\beq
\rho(\{p_i\};p):=\frac{p}{1-p_5}\ketbra{\psi_1}+\frac{1-p}{p_5}\ketbra{\psi_2}
\eeq
equals
\beq
\widehat{\sqrt{\tau_3}}[\rho(\{p_i\};p)]=p \frac{\sqrt{p_1p_4}}{(1-p_5)}\; ,
\eeq
so that we obtain
\beq
\widehat{\sqrt{\tau_3}}[\rho_3]=\sqrt{p_1p_4}\; .
\eeq
These states however have always a non-vanishing concurrence 
$C[\rho_{ij}]=\sqrt{2p_{\!{}_{J_i}}p_{\!{}_{J_j}}}$ and $\vec{J}=(3,2,5,4)$ for non-vanishing 
$p_k$, $k=2,\dots,5$. However, an extended monogamy inequality would be satisfied with 
$\widehat{\sqrt{\tau_3}}^2$ as threetangle, as before.

\subsection{States derived from $\Psi^4_4$}

As we come to the states derived from $\Psi^4_4$ there are two cases to be considered left.
Inserting the weights for the state \eqref{state:4-1} we obtain
\beqa
\ket{\Psi^4_{4;1}}&=&\sqrt{p_1}\ket{0000}+c_2\ket{1100}+
                     c_3\ket{0010}+\nonumber\\
&&\quad \sqrt{p_4}e^{i\eta}\ket{0001}
\eeqa
where the $c_i$ are complex, $|c_i|^2=p_i$, 
and which is normalized if $\sum_{i=1}^4p_i=1$. This state is form-invariant 
under permutations of the first and last two qubits, and hence 
only two different reduced density matrices exist.
They consist of two states which have no three-tangled state in their range.

The second state has the same form-invariance as above.
It is
\beqa
\ket{\Psi^4_{4;2}}&=&c_1\ket{1111}+\sqrt{p_2}\ket{0011}+
                     \sqrt{p_3}e^{i\eta}\ket{0010}+\nonumber\\
&&\quad c_4\ket{0001}
\eeqa
So there are only two essentially different reduced density matrices
\beqa
\tr_1 \ketbra{\Psi^4_{4;2}}&=&p_1 \ketbra{111}+\\
&& \hspace*{-23mm}(\sqrt{p_2}\ket{011}+\sqrt{p_3}e^{i\eta}\ket{010}+c_4\ket{001})({\rm h.c.}) \nonumber\\
\tr_4 \ketbra{\Psi^4_{4;2}}&=& p_3 \ketbra{001}+\\
&& \hspace*{-23mm}(c_1\ket{111}+\sqrt{p_2}\ket{001}+c_4\ket{000})({\rm h.c.}) \nonumber
\eeqa
Whereas the first density matrix has no threetangled state in its range,
the eigenstates of $\tr_4 \ketbra{\Psi^4_{4;2}}$ are
\beqa
\ket{\psi_1}&\propto&\cos \alpha\left(c_1\ket{111}+c_4\ket{000}\right)\\
&&+(\sqrt{p_2}\cos\alpha-\sqrt{p_3}e^{2 i\eta}\sin \alpha)\ket{001}\nonumber\\
\ket{\psi_2}&\propto&\sin \alpha\left(c_1\ket{111}+c_4\ket{000}\right)\\
&&+(\sqrt{p_2}\sin\alpha+\sqrt{p_3}e^{2 i\eta}\cos \alpha)\ket{001}\nonumber
\eeqa
with the condition for orthogonality of the two vectors being
\beqa
\tan (2\alpha)&=&\frac{\sqrt{p_2p_3}}{2(2p_3-1)}\\
\chi&=&-\eta
\eeqa
The weights of the normalized eigenfunctions are
\beqa
P_1=\cos^2\alpha-p_3\cos(2\alpha)-\sqrt{p_1p_3}\sin(2\alpha)\cos(2\eta)\\
P_2=\sin^2\alpha+p_3\cos(2\alpha)+\sqrt{p_1p_3}\sin(2\alpha)\cos(2\eta)
\eeqa
respectively.
This state has a four-fold solution for the vanishing of the threetangle.
They to a zero-polytope which is consisted of a single point in $\CC$
corresponding to a single state whose threetangle vanishes.
The convex roof is known exactly for this situation\cite{AdessoRegula16}.
It is independent of the decomposition of the density matrix,
hence it is a linear function connecting the threetangles of the eigenvectors,
which are
\beqa
\sqrt{\tau_3}[\psi_1]&=&\frac{\cos^2(\alpha)\, \sqrt{p_1 p_4}}{P_1}\\
\sqrt{\tau_3}[\psi_2]&=&\frac{\sin^2(\alpha)\, \sqrt{p_1 p_4}}{P_2}\; .
\eeqa
This results in
\beq
\widehat{\sqrt{\tau_3}}[\tr_4 \ketbra{\Psi^4_{4;2}}]=\sqrt{p_1p_4}\; .
\eeq

The remaining state is
\beqa
\ket{\Psi^4_{4;4}}&=&\sqrt{p_1}\ket{1111}+c_2\ket{1100}+
                     c_3\ket{0010}+\nonumber\\
&&\quad \sqrt{p_4}e^{i\eta}\ket{1110}
\eeqa
where $c_i\in\CC$, $|c_i|^2=p_i$ and with the same condition $\sum_{i=1}^4p_i=1$ for normalization.
This state possesses form-invariance with respect to the first two qubits only.
The reduced density matrices are
\beqa
\tr_1 \ketbra{\Psi^4_{4;4}}&=&p_2 \ketbra{010}+\\
&& \hspace*{-23mm} (\sqrt{p_1}\ket{111}+c_2\ket{100}+\sqrt{p_4}e^{i\eta}\ket{110})
({\rm h.c.}) \nonumber\\
\tr_3 \ketbra{\Psi^4_{4;4}}&=&p_2 \ketbra{010}+\\
&& \hspace*{-23mm} (\sqrt{p_1}\ket{111}+c_3\ket{000}+\sqrt{p_4}e^{i\eta}\ket{110})({\rm h.c.}) \nonumber\\
\tr_4 \ketbra{\Psi^4_{4;4}}&=& p_1 \ketbra{111}+\\
&& \hspace*{-23mm} (c_2\ket{110}+c_3\ket{001}+\sqrt{p_4}e^{i\eta}\ket{111})({\rm h.c.}) \nonumber
\eeqa
The first state has no threetangled state in its whole range; only tracing out 
the third, or the fourth qubit renders a non-zero contribution.

Tracing out the third qubit leads directly to the eigenvectors $\psi_2=\ket{010}$ and
$\psi_1\propto \sqrt{p_1}\ket{111}+c_3\ket{000}+\sqrt{p_4}e^{i\eta}\ket{110}$ 
with corresponding eigenvalues $p_2$ and $1-p_2$, respectively.
As above the convex roof is known exactly to be the linear interpolation
between the eigenstates of $\rho$; hence between zero and
\beq
\sqrt{\tau_3}[\psi_1]=\frac{\sqrt{p_1 p_3}}{1-p_2}\; .
\eeq
Here again, this is trivially seen because the characteristic curves all coincide with a straight 
line which is hence identical with the lowest characteristic curve which is already convex.
Hence, this corresponds to a unique four-fold solution of the zero-polytope\cite{AdessoRegula16}.

Tracing out the fourth qubit gives the eigenstates
\beqa
\ket{\psi_1}&\propto&(\sqrt{p_1}\sin\alpha+\sqrt{p_4}\cos \alpha)\ket{111}\nonumber\\
&&+ e^{i\eta}\cos\alpha(c_2\ket{110}+c_3\ket{001})\\
\ket{\psi_2}&\propto&(\sqrt{p_1}\cos\alpha-\sqrt{p_4}\sin \alpha)\ket{111}\nonumber\\
&&- e^{i\eta}\sin\alpha(c_2\ket{110}+c_3\ket{001})
\eeqa
where 
\beqa
\tan (2\alpha)&=&\frac{2\sqrt{p_1p_4}}{1-2p_1}\; ,\\
\chi&=&\eta\; .
\eeqa
The modulus squared of the eigenstates are
\beqa
P_1&=&\sin^2\alpha +p_1 \cos(2\alpha)-\sin(2\alpha)\sqrt{p_1p_4}\; ,\\
P_2&=&\cos^2\alpha -p_1 \cos(2\alpha)+\sin(2\alpha)\sqrt{p_1p_4}\; .
\eeqa
The convex roof for the threetangle linearly connects 
the tangles of the eigenstates\cite{AdessoRegula16}
\beqa
\sqrt{\tau_3}[\psi_1]&=&\frac{\cos^2(\alpha)\, \sqrt{p_2 p_3}}{P_1}\; ,\\
\sqrt{\tau_3}[\psi_2]&=&\frac{\sin^2(\alpha)\, \sqrt{p_2 p_3}}{P_2}\; .\\
\eeqa

The only non-zero concurrences are $C[\rho_{1,2}]=\sqrt{2p_3p_4}$ and $C[\rho_{3,4}]=\sqrt{2p_1p_2}$,
where $\rho_{i,j}$ is the reduced density matrix of qubits $i$ and $j$.
We state that whenever all the concurrences vanish, also all the threetangles are zero.
We therefor have no perfect analogy to the $W$ states.

Hence, states derived from $\Psi^4_4$ do never lead to a perfect analogue of the $W$-class.
In all cases the derived states do satisfy an extended monogamy relation with 
$\widehat{\sqrt{\tau_3}}^2$ inserted as threetangle.

\section{Conclusions}

In conclusion we have singled out states for four qubits that, 
different from the states from the $W$-class that contain two-tangle,
contain only threetangle which however is globally distributed. 
To this end we have analyzed 
specific four-qubit states which are located in the $SL$ null-cone:
this guarantees that all possible $SL$-invariant four-tangles are zero.
For having states like this, we apply partial spin flips to a c-balanced state\cite{JohanssonO13}.
All states satisfy an extended monogamy relation with $\widehat{\sqrt{\tau_3}}^2$ 
inserted as threetangle. It has however already been excluded that an extension of this 
kind might exist\cite{AdessoOsterlohRegula16}. Since the value of the threetangle will
shrink\cite{OstNineClasses16} (see also Ref.~\cite{NoMonogamy})
with growing $q$ in  $\widehat{\sqrt[q]{\tau_3}}^q$, 
the result will finally be upper bounded by $q=2$.\\
It will be of interest if the various threetangles can be rendered equal.
The latter could be achieved by locally applying $SL$ operations to the states, making use of
$SL$-invariants which scale quadratically in $\psi$ 
(or linearly in $\rho$)~\cite{Viehmann,ViehmannII}.
Also would it be intriguing if such states exist for larger number of qubits $q$ and $n$-site entanglement.
However, for growing number of qubits, the considered reduced density matrices
usually are of higher rank.
It would be nice to see whether translationally (or even permutationally) invariant versions 
of such states will exist and whether it is possible to write such a state for 
arbitrary number of qubits as for the $W$ state.
It is however clear that the permutationally extended version of the state $\ket{\Psi^4_{6;2}}$
with real coefficients will always carry four-tangle unless it becomes a state in the $W$-class
for $p_2=0$.
 
As an interesting byproduct it is demonstrated that the exact convex roof is achieved
for the rank-two case of a homogeneous degree $2m$ polynomial $SL$-invariant measure of entanglement,
if there are states which correspond to 
a maximally $m$-fold degenerate solution in the zero-polytope that can be combined 
with the convexified minimal characteristic curve to give a decomposition of $\rho$.
One has to take the minimum of the results, if more than one such state does exist. 
The threetangle has homogeneous degree 4, hence $m=2$ for this case. 
The minimum over thus constructed decomposition states represents a lower bound to
the $SL$-invariant entanglement measure under consideration; 
it is of course larger than the lowest characteristic curve used 
in Refs.~\cite{Eltschka2012,Siewert2012,EltschkaS12,EltschkaS13}.

We consider it worth to hint towards the alternating signs appearing 
in the monogamy equality of Ref.~\cite{EltschkaPRL15}. 
It could therefore be that a full analogue to the $W$ state may appear only 
for an even number of $n$. In order to test this, one should at least analyze 
corresponding states for five qubits.\\
This alternating sum also appears elsewhere:
for representations of the univeral state inversion\cite{Eltschka17} and 
in the shadow inequalities\cite{Huber17}.
Here merge apparently very different fields as multipartite entanglement and 
quantum error correcting codes. 
Also the Gell-Mann representatives for the operator $\sigma_y$ for qubits
emerging from the representation of the general state inversion\cite{Eltschka17} 
have also appeared before inside the operator with full $SL(d)$ symmetry\cite{Ost15}
that creates the determinant and is used to form the $SL$-invariant analogue to the 
concurrence for qubits.

\section*{Acknowledgements}

We acknowledge fruitful discussions with R. Sch\"utzhold and F. Huber. 
This work was supported by the SFB 1242 of the German Research Foundation (DFG).


\end{document}